\documentclass[twocolumn]{aastex631}
\pdfoutput=1 
\usepackage{appendix}
\usepackage{natbib}
\usepackage{graphicx}
\usepackage{booktabs,threeparttable}
\usepackage{tabularx}
\usepackage{enumitem}
\usepackage{hyperref}
\usepackage{mathtools, amsmath,amstext}
\usepackage{cleveref}
\usepackage[T1]{fontenc}
\usepackage[figure,figure*]{hypcap}

\newcommand{\G}{\mathcal{G}}

\newcommand{\appropto}{\mathrel{\vcenter{\offinterlineskip\halign{\hfil$##$\cr\propto\cr\noalign{\kern2pt}\sim\cr\noalign{\kern-2pt}}}}}

\graphicspath{{./}{figures/}}

\submitjournal{ApJL}
\accepted{May 31, 2024}
\shorttitle{Peas-in-a-Pod Across the Radius Valley}
\shortauthors{Goyal \& Wang}

\begin{document}

\title{Peas-in-a-Pod Across the Radius Valley: \\ Rocky Systems are Less Uniform in Mass but More Uniform in Size and Spacing}

\author[0000-0001-9652-8384]{Armaan V. Goyal}
\affil{Department of Astronomy, Indiana University, Bloomington, IN 47405}

\author[0000-0002-7846-6981]{Songhu Wang}
\affil{Department of Astronomy, Indiana University, Bloomington, IN 47405}

\correspondingauthor{Armaan V. Goyal}
\email{armgoyal@iu.edu}



\begin{abstract}
\noindent 

The ubiquity of ``peas-in-a-pod" architectural patterns and the existence of the radius valley each present a striking population-level trend for planets with $R_{p} \leq 4 R_{\oplus}$ that serves to place powerful constraints on the formation and evolution of these subgiant worlds. As it has yet to be determined whether the strength of this peas-in-a-pod uniformity differs on either side of the radius valley, we separately assess the architectures of systems containing only small ($R_{p} \leq 1.6 R_{\oplus}$), rocky planets from those harboring only intermediate-size ($1.6 R_{\oplus} < R_{p} \leq 4 R_{\oplus}$), volatile-rich worlds to perform a novel direct statistical comparison of intra-system planetary uniformity across compositionally distinct regimes. We find that, compared to their volatile-rich counterparts, rocky systems are less uniform in mass ($2.6\sigma$), but more uniform in size ($4.0\sigma$) and spacing ($3.0\sigma$). We provide further statistical validation for these results, demonstrating that they are not substantially influenced by the presence of mean motion resonances, low-mass host stars, alternative bulk compositional assumptions, sample size effects, or detection biases. We also obtain tentative evidence ($>2 \sigma$ significance) that the enhanced size uniformity of rocky systems is dominated by the presence of super-Earths ($1 R_{\oplus} \leq R_{p} \leq 1.6 R_{\oplus}$), while their enhanced mass diversity is driven by the presence of sub-Earth ($R_{p} < 1 R_{\oplus}$) worlds.
\end{abstract}



\section{Introduction} \label{sec:intro}

Among the several thousands of extrasolar worlds confirmed by NASA's \textit{Kepler} Space Telescope \citep{borucki}, there exists a marked abundance of close-in ($P \lesssim 100$ days) planets smaller than Neptune ($R_{p} \lesssim 4 R_{\oplus} $) that comprise systems with nearly coplanar, almost-circular orbits (\citealt{xie}; \citealt{thompson}; \citealt{weiss_samp}; \citealt{millholland2021}). Aside from embodying the preeminent planetary population in the Galaxy (\citealt{howard_occurence}; \citealt{petigura_occurence}; \citealt{hsu_occurence}; \citealt{he_occurence}), these worlds are also known to exhibit two well-characterized, large-scale trends that serve to place invaluable constraints on the generalized assembly and evolution of multi-planet systems: the ``peas-in-a-pod'' phenomenon and the existence of the radius valley.

The peas-in-a-pod effect describes the highly ubiquitous tendency for subgiant worlds orbiting the same star to be strikingly uniform in their planetary size, mass, and orbital spacing (\citealt{weiss_rad}; \citealt{millholland}; \citealt{wang_2017}; \citealt{goyal}). While its precise origins remain the subject of investigation, this architectural phenomenon has nonetheless elucidated the nature of possible formation channels for multi-planet systems, providing evidence for theoretical frameworks such as global energy optimization of the planetary mass budget (\citealt{adams2}; \citealt{adams}), rapid achievement of the pebble isolation mass (\citealt{lambrechts_iso}; \citealt{johansen_iso}; \citealt{xu_bump}),  convergent migration and energy dissipation within the protoplanetary disk (\citealt{batygin_adams}; \citealt{choksi2}; \citealt{goldberg2}), mass sculpting by postnebular dynamical instabilities (\citealt{izidoro}; \citealt{goldberg}; \citealt{lammers}; \citealt{ghosh}), and construction from a narrow planetesimal ring \citep{batygin_ring}.

The radius valley, also known as the radius gap or Fulton gap, characterizes a significant dearth of planets with $1.5 R_{\oplus} \lesssim  R_{p} \lesssim 2 R_{\oplus}$ that forges a bimodal planetary size distribution likely characterized by small, rocky cores and larger, more volatile-rich worlds (\citealt{owen_2013}; \citealt{fulton_gap}). The volatile component of the latter subpopulation may exist primarily in the form of massive H/He envelopes that are either selectively accreted within gas-poor protoplanetary disks \citep{lee}, or shed at later epochs via photoevaporative processes (e.g. \citealt{owen}; \citealt{mordasini}) or core-powered mass loss (e.g. \citealt{ginzburg}; \citealt{berger_core}). Alternatively, these volatile-rich worlds may instead owe their nature to the bolstered accretion of water beyond the snowline \citep{luque}, where size augmentation is the result of ice sequestered in the core \citep{izidoro2} or a vaporized hydrosphere surrounding the planet \citep{burn}.

Several recent works have considered the possible interplay between these two population-level trends, with notable examples including the provision of the “split peas-in-a-pod” paradigm, for which subgroups of rocky super-Earths and gas-rich sub-Neptunes within a given system each exhibit stronger size uniformity than the system taken as a whole \citep{millholland_split}, and the demonstration of substantial diversity in the core mass fractions of low-mass ($M_{p} \lesssim 10 M_{\oplus}$) rocky planets orbiting the same M-dwarf \citep{rodriguez_martinez}. However, such inquiries have not expressly probed the relationship between peas-in-a-pod architectures and planetary bulk composition. As such, a direct, quantitative comparison of planetary uniformity across the radius valley has yet to be elicited.

We thus perform in this work an explicit statistical comparison of intra-system uniformity between systems containing only small ($R_{p} \leq 1.6 R_{\oplus}$), rocky planets and those with only intermediate-size ($1.6 R_{\oplus} < R_{p} \leq 4 R_{\oplus}$), volatile-rich worlds. Section \ref{sec:gini} introduces our intra-system uniformity metric, while Section \ref{sec:classification} provides justification for our partitioning of planetary composition at $R_{cut} = 1.6 R_{\oplus}$. Section \ref{sec:size} presents our comparative size uniformity analysis for 48 rocky and 118 volatile-rich systems in the California Kepler Survey (CKS, \citealt{weiss_samp}) catalog.  Section \ref{sec:mass} details our mass uniformity analysis, which is predicated upon the robust inference of planetary masses and associated uncertainties for the 48 CKS systems, as well as the consideration of mass measurements for 16 volatile-rich systems from the NASA Exoplanet Archive\footnote{\url{exoplanetarchive.ipac.caltech.edu/docs/data.html}, accessed on December 2, 2023.} (NEA, \citealt{akeson}). We apply in Section \ref{sec:spacing} our spacing uniformity analysis to the 13 CKS rocky and 30 CKS volatile-rich systems with $N_{p} \geq 3$.
We then evaluate in Section \ref{sec:sub_earth} the isolated behavior of the sub-Earth ($R_{p} < 1 R_{\oplus}$) and super-Earth ($1 R_{\oplus} \leq R_{p} \leq 1.6 R_{\oplus}$) subpopulations within our CKS rocky sample. Finally, we briefly explore the possible astrophysical implications of our findings and discuss the context of our results with regard to the inner solar system in Section \ref{sec:discussion}. We consider the potential effects of various confounding factors and statistical biases on our primary results in Appendix \ref{sec:bias}.

\begin{figure*}
  \includegraphics[width=\textwidth]{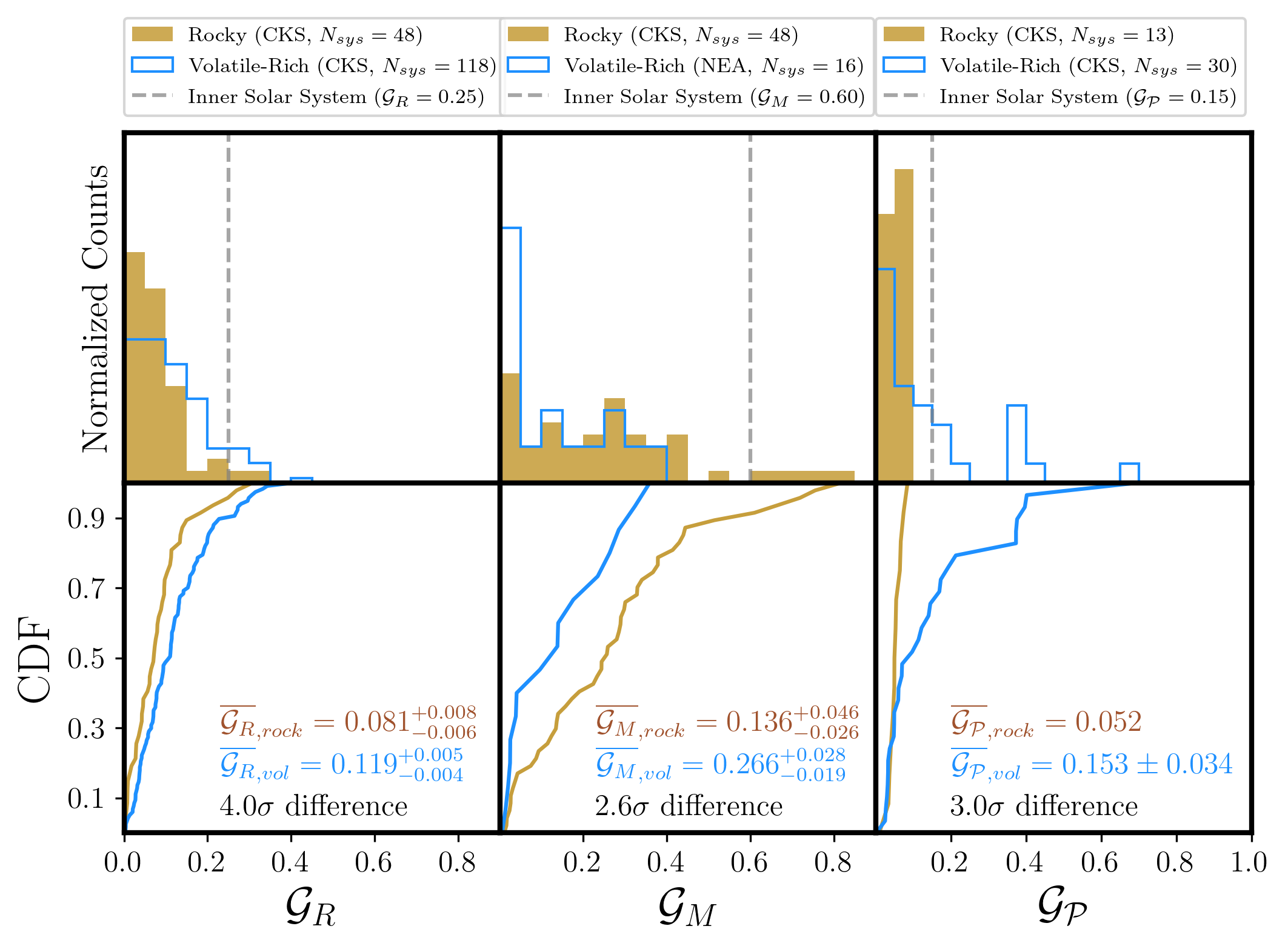}
  \caption{Left: Rocky systems generally favor configurations more uniform in planetary size, less uniform in planetary mass and more uniform in planetary spacing than their volatile-rich counterparts. Left: The distribution of $\mathcal{G}_{R}$ values within our CKS rocky sample (gold) is 73\% and 92\% contained within the respective high-uniformity regimes of $\mathcal{G}_{M} \leq 0.1$ and  $\mathcal{G}_{M} \leq 0.2$ , compared to only 49\% and 84\% of the CKS volatile-rich sample distribution (blue). Following the uncertainty propogation procedure described in Section \ref{sec:size_results}, we find that rocky systems exhibit enhanced intra-system size uniformity compared to volatile-rich systems at the level of $4.04 \sigma$ significance. Middle: 27\% of the CKS rocky mass Gini distribution lies above the maximum value for the NEA volatile-rich systems  ($\mathcal{G}_{M} = 0.36$). As described in Section \ref{sec:mass_result}, rocky systems exhibit greater planetary mass diversity compared to volatile-rich systems at the level of $2.61 \sigma$ significance. Right: The distribution of $\mathcal{G}_\mathcal{P}$ values within our rocky sample (gold) is completely contained within $\mathcal{G}_\mathcal{P} \leq 0.1$, a value above which 53\% of the volatile-rich population resides. As described in Section \ref{sec:spacing_res}, rocky systems exhibit enhanced intra-system spacing uniformity compared to volatile-rich systems at the level of $3.0 \sigma$ confidence. Gray dashed lines indicate size ($\mathcal{G}_{M} = 0.25$), mass ($\mathcal{G}_{M} = 0.60$) and spacing ($\mathcal{G}_{\mathcal{P}} = 0.15$) Gini values for the four terrestrial planets in our solar system. We further discuss the contextualization of the inner solar system in Section \ref{sec:theory}}
  \label{gini_hists}
\end{figure*}
    
\section{Uniformity Metric: The Gini Index} \label{sec:gini}

In accord with the analyses of \citet{goyal} and \citet{goyal2}, we adopt the adjusted Gini index \citep{deltas_2003}, a common economic measure of wealth or income inequality in a given population, as our primary metric for the assessment of intra-system mass and spacing uniformity. The adjusted Gini index can be expressed mathematically as:

\begin{equation}\label{eq1}
    \mathcal{G} = \left(\frac{N}{N-1}\right)G,
\end{equation}

where $G$ is the standard Gini index \citep{gini}, given for a data vector $x$ with size $N$ by:

\begin{equation}
    G = \frac{1}{2N^{2}\overline{x}}\sum_{i=1}^{N} \sum_{j=1}^{N} |x_{i}-x_{j}|.
\end{equation}

$\mathcal{G}$ is normalized to unity by construction, such that an data vector with uniform entries would yield $\mathcal{G} = 0$, while a maximally diverse set of entries would yield $\mathcal{G} = 1$. The additional $N/(N-1)$ prefactor in $\mathcal{G}$ acts as a corrective term for an inherent downward bias in $G$ for $N \lesssim 10$ \citep{deltas_2003}, and has been shown to successfully is stabilize this effect to within $\sim2$\% when applied to CKS planetary systems with $N_{p} \in [2, 5]$ \citep{goyal}. 

All references to the Gini index in this work shall refer only to the adjusted Gini index $\mathcal{G}$ given by Eq. \ref{eq1}. We shall note also that although alternative, conceptually similar metrics have been utilized in other inquiries for the purposes of quantifying intra-system mass and spacing uniformity, such as total pariwise logarithmic distance \citep{millholland} or unit-normalized mass partitioning and gap complexity \citep{gilbert}, we maintain our usage of the Gini index due to its relative computational simplicity, as well as the abundance of existing literature regarding its applications and intricacies.

\begin{figure}
  \includegraphics[width=0.45\textwidth]{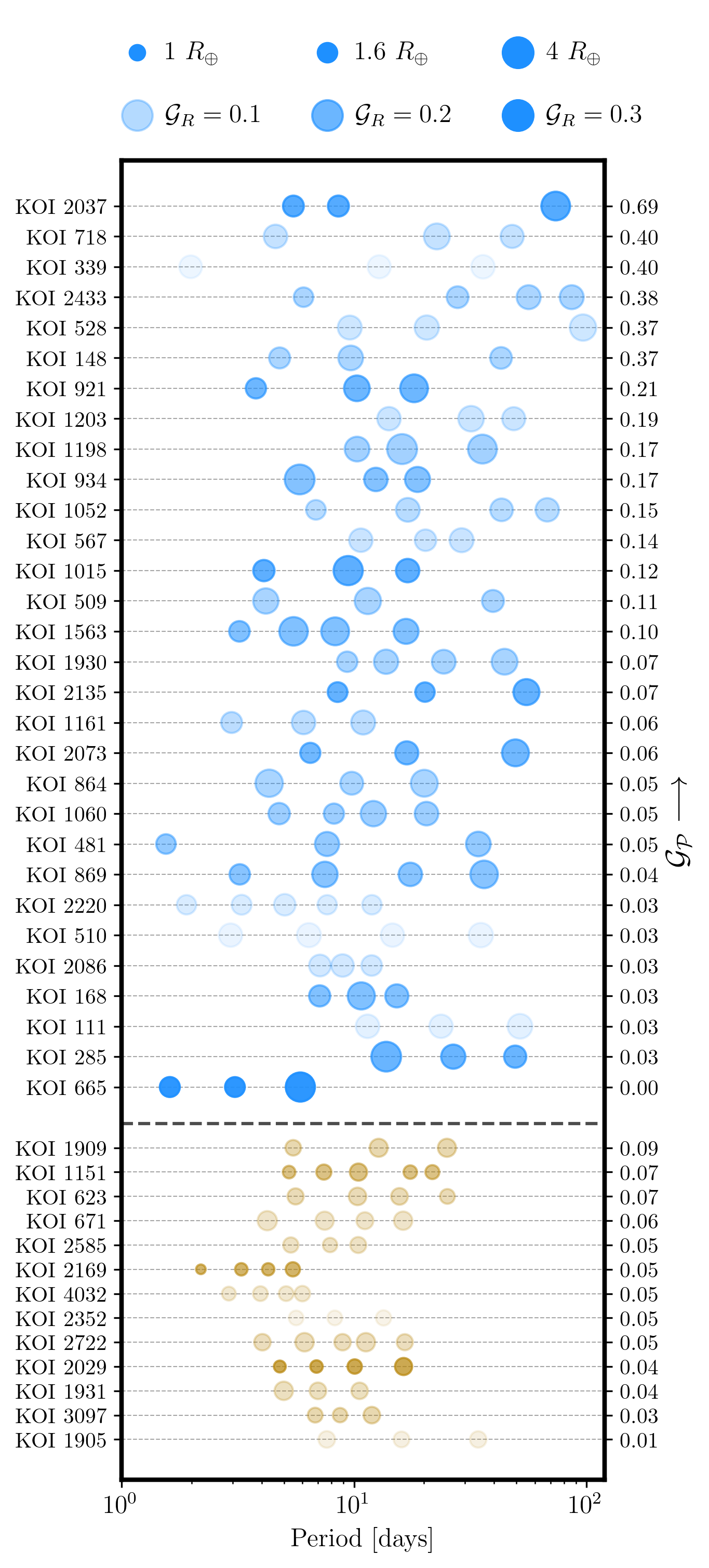}
    \caption{Orbital architectures of the 13 high-multiplicity ($N \geq 3$) systems in our CKS rocky sample (gold), and the 30 high-multiplicity systems in our CKS volatile-rich sample (blue). This high-$N$ restriction is imposed to avoid concurrent illustration of over 150 systems, the full CKS rocky and volatile-rich samples are described in Section \ref{sec:size_samp}. Marker sizes and transparency respectively correspond to planetary size and system-wide size uniformity (greater opacity indicates higher $\mathcal{G}_{R}$). The generally lighter shading of the rocky systems indicates a qualitative preference for enhanced size uniformity, which we verify with greater statistical rigor in Section \ref{sec:size_results}. Within both samples, systems are also plotted in vertical order of decreasing spacing uniformity (increasing $\mathcal{G}_{\mathcal{P}}$). We observe that all rocky systems have  $\mathcal{G}_{\mathcal{P}} < 0.1$ while approximately half of the volatile-rich systems reside above this regime, indicating comparatively greater intra-system spacing uniformity for the former population.}
  \label{sys_arch}
\end{figure}

\section{Compositional Classification Scheme} \label{sec:classification}
Empirical modeling of the mass-radius ($M$-$R$) relation for subgiant worlds ($R_{p} \lesssim 4 R_{\oplus}$, $M_{p} \lesssim 100 M_{\oplus}$) favors a transition from small, rocky spheres of approximately constant density ($R_{p} \sim M_{p} ^{0.33}$) to intermediate-size planets with steeper radius inflation ($R_{p} \sim M_{p}^{0.6}$) owing to the enhanced contribution of volatiles (\citealt{chen_kipping}; \citealt{bashi}; \citealt{zeng_2019}; \citealt{otegi_2020}; \citealt{edmondson}; \citealt{baron}). \citet{weiss2014} posited that these two regimes of the $M$-$R$ parameter space may be separated by a threshold radius $R_{cut} \approx 1.5 R_{\oplus}$, while \citet{rogers} determined via a hierarchical  Bayesian framework that this boundary may instead lie closer to $R_{cut} \approx 1.6 R_{\oplus}$, a value that has since been corroborated by recent empirical fitting of expanded planetary catalogs (\citealt{lozovsky}; \citealt{baron}). Furthermore, it has been shown that extant precise ($\delta M_{p}/M_{p} \lesssim 0.3$) mass measurements for rocky worlds with $R_{cut} \leq 1.6 R_{\oplus}$ yield exceptionally close agreement (\citealt{rodriguez_martinez}; \citealt{rubenzahl}) with the semi-empirical $M$-$R$ relation for Earth-composition (30\% Fe - 70\% MgSiO$_{3}$) planets presented by \citet{zeng}. As such, we adopt hereafter this $R_{cut} = 1.6 R_{\oplus}$ threshold as the basis for classification of rocky and volatile-rich worlds throughout this analysis.

\begin{table*}
\centering
\caption{Summary of Size Uniformity Tests: CKS Rocky ($N_{sys} = 48$) vs. CKS Volatile-Rich ($N_{sys} = 118$)}
\begin{tabular}{c |c |c|c| c} 
 \multicolumn{5}{c}{} \\
 Description & $N_{sys, rock}$ & $N_{sys, vol}$ & Significance Result & Ref. in Text\\
 \hline \hline

  Primary Size Uniformity Comparison & 48 & 118 & $4.04\sigma$ & Sec. \ref{sec:size_results}\\

No Compositional Dichotomy (Assess Sample Size Effects) & 48 & 118 & $0.00\sigma\pm1.66\sigma$ & Sec. \ref{sec:samp_size}\\

Remove Systems with any $R_{cut} = 1.6 R_{\oplus} \in [R_{p} - \delta R_{p}, R_{p} + \delta R_{p}]$ & 42 & 96 & $1.76\sigma$ & Sec. \ref{sec:overlap}\\ 

Remove Near-Resonant Systems & 37 & 102 & $3.54\sigma$ & Sec. \ref{sec:near_res}\\ 

    FGK Main-Sequence Hosts Only & 48 & 103 & $4.04\sigma$ & Sec. \ref{sec:fgk}\\ 

    Remove Systems with Possible Undetected Planet within $P \leq 100$ days & 23 & 110 & $3.68\sigma$ & Sec. \ref{sec:missing_plan}\\\hline
 
 \end{tabular}
 \label{size_tab}
\end{table*}

\section{Planetary Size Uniformity}
\label{sec:size}
\subsection{Selection of Rocky \& Volatile-Rich Samples from the CKS} \label{sec:size_samp}
In order to promote the greatest degree of statistical integrity with regard to a comparative, population-level analysis, we consider the 892 planets across 349 multi-planet systems from the CKS \citep{weiss_samp}, for which the usage of consistent instrumentation and survey construction across their photometric, spectroscopic, and astrometric measurements yields a homogeneous catalog of high-fidelity planetary radii refined to $\sim 10 \%$ precision (\citealt{johnson}; \citealt{fulton}). 

We thus construct our CKS rocky sample by selecting from these 349 multi-planet systems those wherein all constituent planets have orbital periods and radii respectively satisfying $P \leq 100$ days and $R_{p} \leq 1.6 R_{\oplus}$, obtaining 48 systems containing a total of 118 planets. We construct our CKS volatile-rich sample in an analogous manner by altering the radius constraint to $1.6 R_{\oplus} \leq R_{p} \leq 4 R_{\oplus}$, obtaining 118 systems containing a total of 275 planets. We calculate the size Gini index $\mathcal{G}_{R}$ of each system in both samples and compare the associated distributions in the left panel of Figure \ref{gini_hists}. We find for the CKS rocky sample that 73\% and 92\% of Gini values respectively lie within $\mathcal{G}_{R} \leq 0.1$ and $\mathcal{G}_{R} \leq 0.2$, while only 49\% and 84\% of the CKS volatile-rich systems occupy the same high-uniformity regime. We find also that the 90th percentile of the rocky $\mathcal{G}_{R}$ distribution respectively corresponds to a value of 0.16, a region above which 26\% of the volatile-rich systems reside. We also plot system architectures for the $N_{p} \geq 3$ subset of either sample (restriction imposed for convenience of illustration as well as to promote correspondence with Section \ref{sec:spacing}), and observe the same relative proclivity of the CKS rocky systems for occupying configurations with low values of $\mathcal{G}_{R}$. Rocky systems thus exhibit a qualitative preference for greater planetary size uniformity compared to their volatile-rich analogs, a trend which we shall assess more rigorously in the proceeding section.

\vspace{3mm}

\subsection{Size Uniformity Analysis and Results}
\label{sec:size_results}
Our primary statistical comparison of intra-system size uniformity is based upon the propagation of measurement uncertainties in planetary radius ($\delta R_{p}$) to analogous bounds on the respective mean size Gini values ($\overline{\mathcal{G}_{R}}$) for the CKS rocky and volatile-rich samples. Within each system, we perform $10^{4}$ iterations of a uniform random draw of each planetary radius from its constituent $1\sigma$ uncertainty distribution $R_{p} \pm \delta R_{p}$. We calculate the system Gini from these redrawn radii at each iteration to obtain a distribution of $\mathcal{G}_{R}$ values for the system itself, the bounds of which are defined as $\mathcal{G}_{R} \pm \delta \mathcal{G}_{R}$. Across a sample with $N_{sys}$ systems, we calculate the mean size Gini $\overline{\mathcal{G}_{R}}$, and obtain the associated $1\sigma$ uncertainty values via propagation of the size Gini uncertainties from the individual systems, with $\delta \overline{\mathcal{G}_{R}} = \sqrt{\sum (\delta \mathcal{G}_{R})^{2}}/N_{sys}$. Following this methodology, we obtain for our CKS rocky and volatile-rich distributions respective values of $\overline{\mathcal{G}_{R, rock}} = 0.081_{-0.006}^{+0.008}$ and $\overline{\mathcal{G}_{R, vol}} = 0.119_{-0.004}^{+0.005}$, indicative of enhanced size uniformity for the CKS rocky systems at the level of $4.04\sigma$ significance.

We perform several additional tests to validate this result in Appendix \ref{sec:bias}, considering the influence of sample size effects (Section \ref{sec:samp_size}), planets that overlap the $R_{cut} = 1.6 R_{\oplus}$ boundary (Section \ref{sec:overlap}), the presence of high-uniformity configurations near mean motion resonance (Section \ref{sec:near_res}), non-FGK host stars (Section \ref{sec:fgk}), and the possibility of missing planets (Section \ref{sec:missing_plan}). We list the results of these ancillary tests, as well as our primary size uniformity result, in Table \ref{size_tab}.

\begin{table*}
\centering
\caption{Summary of Mass Uniformity Tests: CKS Rocky ($N_{sys} = 48$, Inferred Mass) vs. NEA Volatile-Rich ($N_{sys} = 16$, Measured Mass)}
\begin{tabular}{c |c |c|c| c} 
 \multicolumn{5}{c}{} \\
 Description & $N_{sys, rock}$ & $N_{sys, vol}$ & Significance Result & Ref. in Text\\
 \hline \hline

  Primary Mass Uniformity Comparison & 48 & 16 & $2.61\sigma$ & Sec. \ref{sec:mass_result}\\

Iron-Rich $M$-$R$ Relation for $R_{p} \leq 1.6 R_{\oplus} $ & 48 & 16 & $2.53\sigma$ & Sec. \ref{sec:comp_assume}\\

Pure $M$-$R$ Relation for $R_{p} \leq 1.6 R_{\oplus} $ & 48 & 16 & $2.51\sigma$ & Sec. \ref{sec:comp_assume}\\

Random $M$-$R$ Relation (Earth, Rock, Iron) for $R_{p} \leq 1.6 R_{\oplus} $ & 48 & 16 & $2.66\sigma$ & Sec. \ref{sec:comp_assume}\\

No Compositional Dichotomy (Assess Sample Size Effects) $\leq 1.6 R_{\oplus} $ & 48 & 16 & $0.03\sigma\pm1.20\sigma$ & Sec. \ref{sec:samp_size}\\

Remove Systems with any $R_{cut} = 1.6 R_{\oplus} \in [R_{p} - \delta R_{p}, R_{p} + \delta R_{p}]$ & 42 & 13 & $2.41\sigma$ & Sec. \ref{sec:overlap}\\ 

Remove Near-Resonant Systems & 37 & 12 & $2.67\sigma$ & Sec. \ref{sec:near_res}\\ 

    FGK Main-Sequence Hosts Only & 48 & 9 & $2.60\sigma$ & Sec. \ref{sec:fgk}\\ 

Only Mass Measurements for $R_{p} > 1.6 R_{\oplus}$& 48 & 14 & $2.53\sigma$ & Sec. \ref{sec:ttv_rv}\\\hline
 
 \end{tabular}
 \label{mass_tab}
\end{table*}

\section{Planetary Mass Uniformity}
\label{sec:mass}
\subsection{Inference of Planetary Masses for CKS Rocky Systems} \label{sec:mass_inference}
Across the 166 collective systems within our CKS rocky and volatile-rich samples, only one system (KOI-321, \citealt{marcy}) harbors extant mass measurements for each of its constituent planets, indicating that these CKS samples lack the requisite associated planetary mass data for a population-level uniformity analysis. However, this dearth may be alleviated in part by the substantial efficacy with which the $M$-$R$ bivariate behavior of small, rocky ($R_{p} \leq 1.6 R_{\oplus}$) worlds may be modeled (see Fig. 5 of \citealt{rodriguez_martinez}; Fig. 8 of \citealt{rubenzahl}) by the semi-empirical $M$-$R$ relation for Earth-like planets presented by \citet{zeng}. We therefore utilize this \citet{zeng} Earth-composition model to robustly infer planetary masses $M_{p}$ for the 118 objects (across 48 systems) within our CKS rocky sample (Figure \ref{MR}, gold curve and points). Since this $M$-$R$ relation is itself monotonic, we also determine $1\sigma$ uncertainties $\delta M_{p}$ on the inferred planetary masses based on direct interpolation of the CKS-derived $1\sigma$ radius interval $R_{p} \pm \delta R_{p}$ associated with each object (Figure \ref{MR}).

\subsection{Construction of Volatile-Rich Sample from NASA Exoplanet Archive} \label{sec:nea_vol}
In contrast to their rocky counterparts, volatile-rich worlds do not exhibit any tight, universally applicable correspondence between their planetary masses and radii, due primarily to the greater uncertainty imposed by their individual compositions (\citealt{hatzes}; \citealt{otegi_2020}; \citealt{jontoff-hutter}), as well as their propensity for size inflation due to small changes in internal or surface-driven processes (e.g. \citealt{kite}). As such, their planetary masses may not be inferred from any global $M$-$R$ relation, and must be characterized exclusively by precise mass measurement. We thus construct our volatile-rich sample directly from the NEA \footnote{We shall briefly note that since the data provided by the NEA are the aggregate of planetary parameters derived from independent surveys, instruments, pipelines, and analysis techniques, any sample of planetary systems taken therefrom will inherently exhibit statistical heterogeneities that may propagate forward through any subsequent analysis. These heterogeneities may themselves present as complex biases resulting from inherent differences in systematics and heuristic determination of planetary parameters, as well as nonstandardized determination of errors across the entire sample. While a detailed exploration of such heterogeneity-driven biases and their associated statistical effects are beyond the scope of this work, we nonetheless acknowledge here the presence of these caveats as they pertain to our own sample and analysis.}, which we query for systems based on the following selection criteria: 

\begin{enumerate}
    \item All constituent planets exhibit non-grazing ($b \leq 0.9$) transits.
    \item All constituent planets have radii satisfying $1.6 R_{\oplus} = R_{cut}< R_{p} \leq 4 R_{\oplus}$
    \item All constituent planets have associated mass measurements (such that determinations of intra-system uniformity will not be biased by incomplete system mass characterization).
    \item All constituent planetary mass measurements correspond to true mass estimates (i.e. values corresponding to $M_{p} \sin(i)$ or upper/lower limits are not considered) derived from either radial velocity (RV) or transit timing variation (TTV) data.
    \item Any nominal TTV mass measurements adopted from \citet{hadden_2014} or \citet{hadden} originate from the ``low-$e$” subsample of the former work or the ``robust” subsample of the latter.
    \item All constituent mass measurements have fractional uncertainty values satisfying $\delta M_{p}/M_{p} \leq 1$.
    \item All planets have orbital periods of $P \leq 100$ days.
    
\end{enumerate}

\begin{figure*}
  \includegraphics[width=\textwidth]{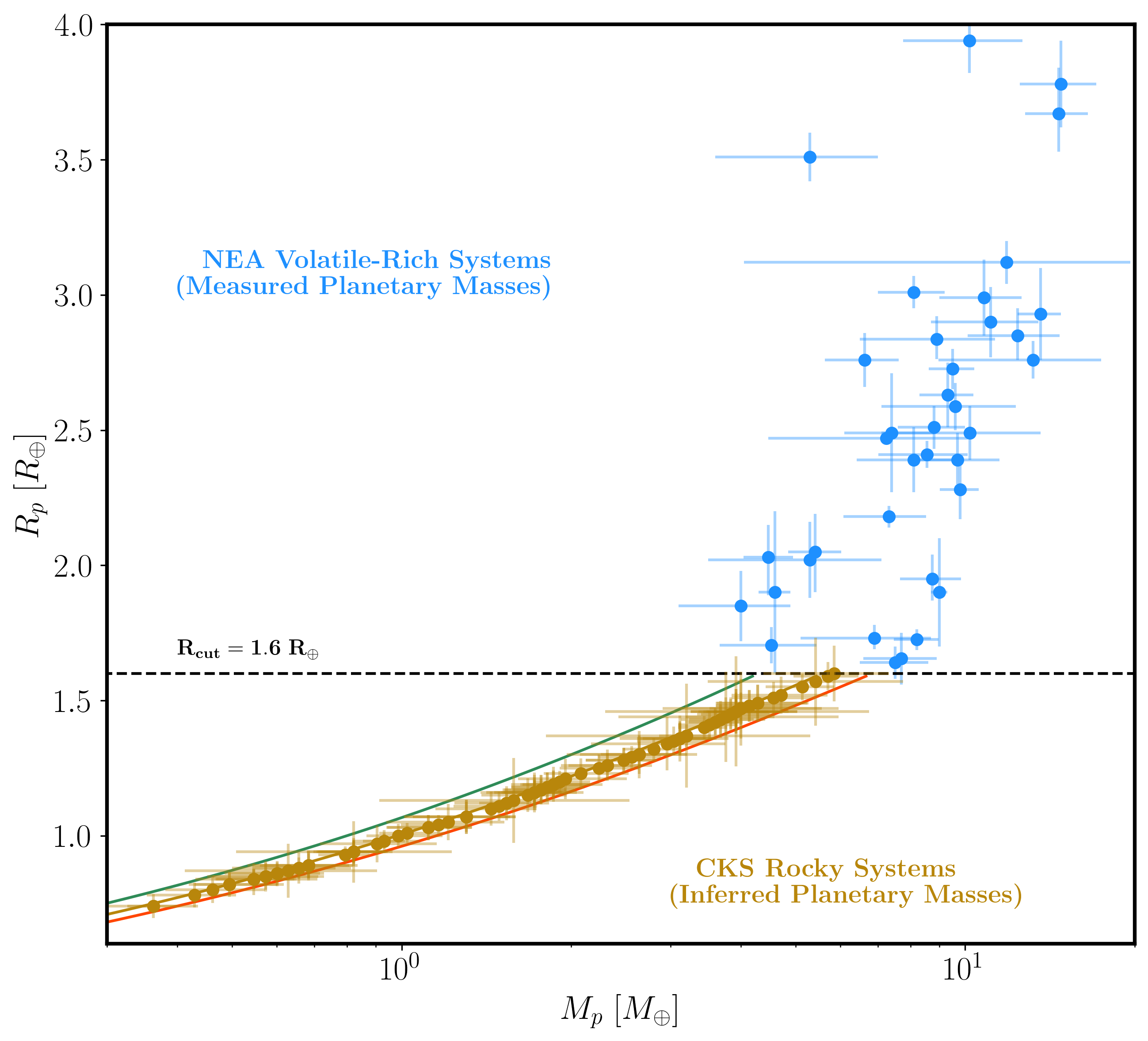}
  \caption{Left: Compositional dichotomy between small, rocky planets and intermediate-size, volatile-rich planets, as defined by a transition radius $R_{cut} = 1.6 R_{\oplus}$ (Section \ref{sec:classification}). In accord with the findings of \citet{rodriguez_martinez} and \citet{rubenzahl}, we leverage the tight correspondence of observed rocky worlds with the semi-empirical Earth-composition (30\%Fe – 70\% MgSiO$_{3}$) $M$-$R$ relation (gold curve) presented by \citet{zeng} to robustly infer planetary masses $M_{p}$ and associated $1\sigma$ mass uncertainties $\delta M_{p}$ for 118 small ($R_{p} \leq 1.6 R_{\oplus}$) worlds (gold points) across 48 CKS systems (\citealt{weiss_samp}, see Section \ref{sec:mass_inference} of this work). Our NEA volatile-rich sample (blue points, see Section \ref{sec:nea_vol} of this work) exhibits qualitatively stronger clustering in planetary mass than the CKS rocky worlds, indicative of a greater degree of planetary mass diversity for rocky systems, which we assess rigorously in Section \ref{sec:mass_result}. We explore in Section \ref{sec:comp_assume} the usage of iron-rich (red curve, 50\% Fe – 50\% MgSiO$_{3}$) and pure rock (green curve) relations from \citet{zeng} as alternative compositional models for the mass inference of our CKS rocky planets, finding that the enhanced mass diversity of rocky systems increases with the simultaneous consideration of multiple possible bulk planetary compositions as opposed to an Earth-like $M$-$R$ profile alone.}
  \label{MR}
\end{figure*}

We obtain from our query of the NEA 16 systems, containing a total of 37 intermediate-size planets, that satisfy the aforementioned criteria. We place these 37 planets within the $M$-$R$ parameter space in Figure \ref{MR}. We note that none of these 16 NEA volatile-rich systems overlap with 118 CKS volatile-rich systems treated in Section \ref{sec:size}.

As in Section \ref{sec:size}, we compare in the central panel of Figure \ref{gini_hists} the mass Gini $(\mathcal{G}_{M})$ distributions of the CKS rocky sample (following the mass inference procedure thus described) and the NEA volatile-rich sample, finding that 25\% and 42\% of the CKS rocky values respectively lie within $\mathcal{G}_{M} \leq 0.1$ and $\mathcal{G}_{M} \leq 0.2$ compared to 50\% and 69\% of the NEA volatile-rich sample, and that 27\% of the former distribution lies above the maximum value of latter ($\mathcal{G}_{M} = 0.36$). We thus observe, in possible contrast to our findings for size uniformity, that rocky systems may qualitatively favor configurations with a greater diversity in planetary mass.

\subsection{Mass Uniformity Analysis and Results}
\label{sec:mass_result}
Having determined $1\sigma$ uncertainties $\delta M_{p}$ for the inferred planetary masses across our 48 CKS rocky systems, and with extant measurement uncertainties in planetary mass for our 16 NEA volatile-rich systems, we perform our intra-system mass uniformity analysis with the same uncertainty propagation methodology described in Section \ref{sec:size_results}. We obtain for our CKS rocky and volatile-rich distributions respective mean Gini values of $\overline{\mathcal{G}_{M, rock}} = 0.266_{-0.019}^{+0.028}$ and $\overline{\mathcal{G}_{M, vol}} = 0.136_{-0.026}^{+0.046}$, confirming the presence of increased mass diversity for the CKS rocky systems at the level of $2.61\sigma$ significance.

As in Section \ref{sec:size_results}, we perform several additional tests to validate our mass uniformity result in Appendix \ref{sec:bias}. We demonstrate in Section \ref{sec:comp_assume} that the inference of individual planetary masses from multiple distinct compositional curves (Fig. \ref{MR}) instead of from single $M$-$R$ relation further increases the level of mass diversity observed for the CKS rocky sample, indicating that our $2.61\sigma$ result is likely a lower limit on the true level of discrepancy between rocky and volatile-rich systems. We also repeat many of the same supplementary tests as described in Section \ref{sec:size_results}, as well as an assessments of the influence of volatile-rich systems with any TTV mass measurements (Section \ref{sec:ttv_rv}). We provide a summary of these results in Table \ref{mass_tab}.

\begin{table*}
\centering
\caption{Summary of Spacing Uniformity Tests: High-$N$ CKS Rocky ($N_{sys} = 13$) vs. High-$N$ CKS Volatile-Rich ($N_{sys} = 30$)}
\begin{tabular}{c |c |c|c| c} 
 \multicolumn{5}{c}{} \\
 Description & $N_{sys, rock}$ & $N_{sys, vol}$ & Significance Result & Ref. in Text\\
 \hline \hline

  Primary Spacing Uniformity Comparison & 13 & 30 & $3.01\sigma$ & Sec. \ref{sec:spacing_res}\\

No Compositional Dichotomy (Assess Sample Size Effects) & 13 & 30 & $-0.04\sigma\pm1.71\sigma$ & Sec. \ref{sec:samp_size}\\

Remove Systems with any $R_{cut} = 1.6 R_{\oplus} \in [R_{p} - \delta R_{p}, R_{p} + \delta R_{p}]$ & 11 & 24 & $2.88\sigma$ & Sec. \ref{sec:overlap}\\ 

    FGK Main-Sequence Hosts Only & 13 & 29 & $3.11\sigma$ & Sec. \ref{sec:fgk}\\ 

    Remove Systems with Possible Undetected Planet within $P \leq 100$ days & 8 & 29 & $2.01\sigma$ & Sec. \ref{sec:missing_plan}\\
 
 \end{tabular}
 \label{spacing_tab}
\end{table*}

\section{Planetary Spacing Uniformity}
\label{sec:spacing}
\subsection{High-Multiplicity Subsets of CKS Rocky \& Volatile-Rich Samples} \label{sec:spacing_samp}
Given that our intended statistical evaluations of planetary spacing are agnostic to planetary mass, and in order to further promote of the greatest degree of statistical homogeneity within our samples and the analyses based thereupon, our investigation of intra-system spacing uniformity exclusively considers the CKS rocky and volatile-rich samples utilized for our size uniformity analysis in Section \ref{sec:size}. However, since uniformity in interplanetary spacing may only be assessed for systems with at least two distinct planetary pairs, we limit our consideration to the subset of ``high-$N$'' systems in either sample which contain at least 3 planets ($N_{p} \geq 3$). In this regard, we respectively obtain for our high-$N$ CKS rocky and volatile-rich samples 13 systems containing 48 planets and 30 systems containing 99 planets. We utilize the period ratios of neighboring planetary pairs ($\mathcal{P} = P_{i+1}/P_{i}$) as our metric for interplanetary spacing within each system, and subsequently calculate the spacing Gini $\mathcal{G}_{\mathcal{P}}$ for these 43 combined systems and illustrate their orbital architectures in Figure \ref{sys_arch}, where we observe a qualitative preference for rocky systems to occupy more uniformly-spaced architectures. This preference is corroborated by the right panel of Figure \ref{gini_hists}, where the high-$N$ CKS rocky distribution is fully contained within $\mathcal{G}_{\mathcal{P}} \leq 0.1$ while 53\% of the high-$N$ CKS volatile systems reside above this mark.

\subsection{Spacing Uniformity Analysis and Results} \label{sec:spacing_res}
Given that all \textit{Kepler} planetary orbital periods are precisely determined from transit photometry such that associated fractional uncertainties are themselves generally of order $\delta P/P \leq 10^{-5}$ \citep{lissauer_per}, we do not adopt for our spacing uniformity analysis the uncertainty propagation procedure utilized in Sections \ref{sec:size} and \ref{sec:mass}. Instead, we shall consider instead the following null hypothesis testing framework put forth by \citet{goyal2}: we adopt the null hypothesis that the high-$N$ CKS rocky and volatile-rich samples exhibit no difference in the degree of their intra-system spacing uniformity, and thus compare the mean spacing Gini $\overline{\mathcal{G}_{\mathcal{P}}}$ of the 13 high-$N$ CKS rocky systems to a control distribution of $\overline{\mathcal{G}_{\mathcal{P}}} \pm \delta \overline{\mathcal{G}_{\mathcal{P}}}$  calculated for $10^{5}$ subsamples of 13 systems drawn randomly (without replacement) from the 30 high-$N$ CKS volatile-rich systems. From this procedure, we obtain for our 13 high-$N$ CKS rocky systems and our 30 high-$N$ CKS volatile-rich systems respective values of $\overline{\mathcal{G}_{\mathcal{P},rock}} = 0.052$ and $\overline{\mathcal{G}_{\mathcal{P}, vol}} = 0.153 \pm 0.034$, indicating a $3.01 \sigma$ discrepancy corresponding to enhanced spacing uniformity for rocky systems.

We provide a greater exploration of this spacing uniformity result in Appendix \ref{sec:bias} using the same additional tests outlined at the end of Section \ref{sec:size_results}. Table \ref{spacing_tab} provides a summary of these results.

\section{Influence of Sub-Earths within Rocky Systems}
\label{sec:sub_earth}
Within the general landscape of small, rocky worlds ($R_{p} \leq 1.6 R_{\oplus}$), observed by Kepler, there has been recent evidence to suggest the presence of a transition near $R_{p} \approx 1 R_{\oplus}$ from super-Earths to a unique population of “sub-Earths” ($R_{p} < 1 R_{\oplus}$) that may harbor a greater occurrence rate and distinct dynamical histories compared to larger terrestrial worlds (\citealt{hsu_occurence}; \citealt{qian}). In stark contrast to the combined population of super-Earths and sub-Neptunes ($1 R_{\oplus} \leq R_{p} \leq 4 R_{\oplus}$) and the well-characterized size bimodality that lies therein (\citealt{owen_2013}; \citealt{fulton_gap}), sub-Earths are best described by a power-law size-frequency distribution \citep{qian} analogous to those observed for asteroids (e.g. \citealt{johansen}; \citealt{trilling}) and Kuiper Belt objects (e.g. \citealt{fraser}; \citealt{morbidelli_kbo}). This quality, in conjunction with the tendency for such objects to coexist with at least one additional transiting companion, may suggest that sub-Earths are not the outcomes of planet formation within a primordial gaseous disk, but are instead the late-stage products of postnebular debris disks \citep{qian}. Given the high level of size and mass diversity expected for objects with such a dynamical history, we shall attempt to investigate the influence of such sub-Earths on the comparative uniformity of rocky and volatile-rich systems.

\begin{figure*}
  \includegraphics[width=\textwidth]{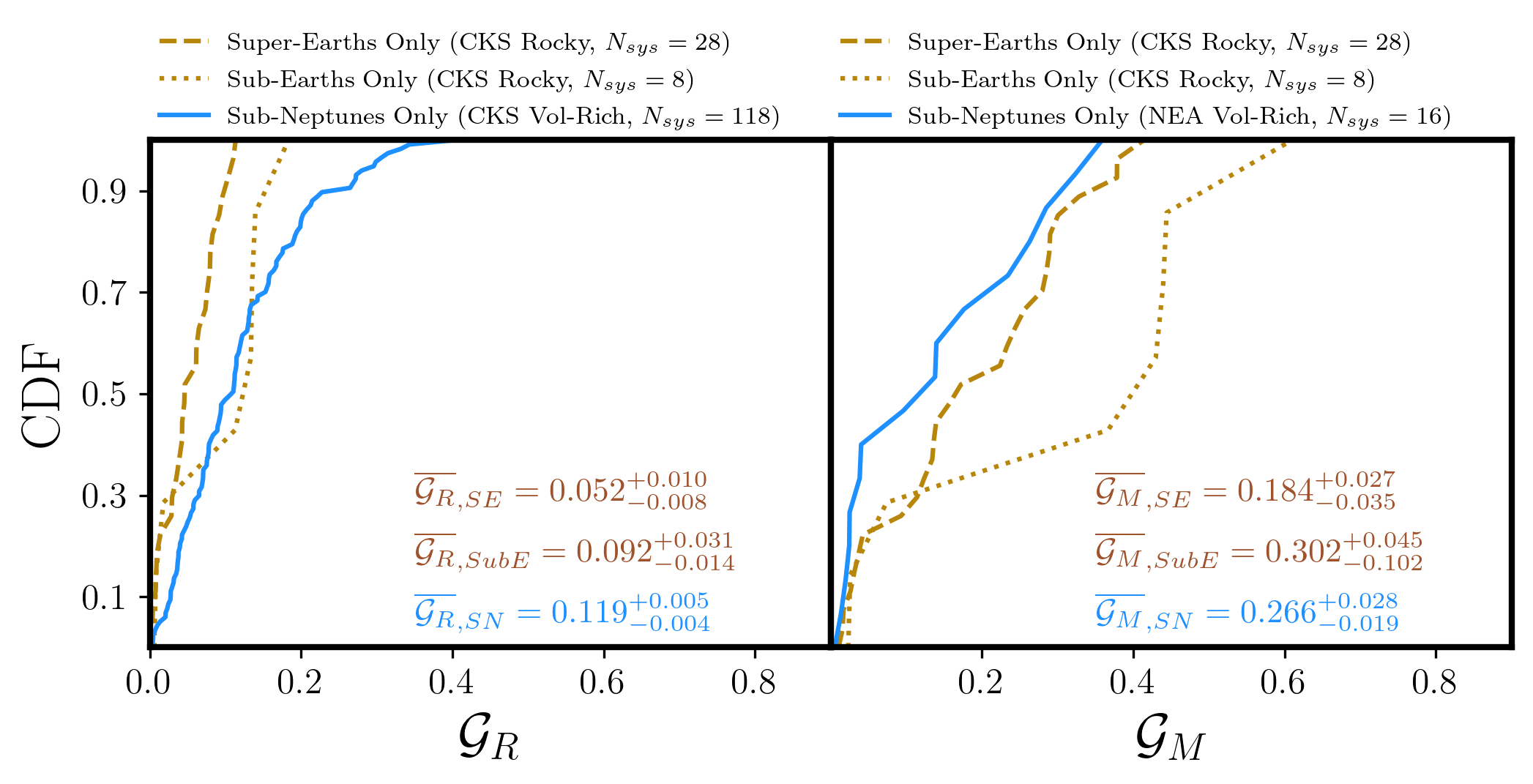}
  \caption{Enhanced size uniformity and mass diversity of the CKS rocky systems are driven by the respective presence of super-Earths and sub-Earths. Left: Subject to the testing procedure outlined in Section \ref{sec:size_results}, CKS rocky systems containing only sub-Earths (SubEs, $R_{p} < 1 R_{\oplus}$) are more diverse in planetary size than systems containing only super-Earths (SEs, $1 R_{\oplus} \leq R_{p} \leq 1.6 R_{\oplus}$) with $\sim2.0\sigma$ significance. Taking our 118 CKS volatile-rich systems as a representative sample of systems containing only sub-Neptunes (SNs), we find that the size uniformity of SubE-only and SN-only systems are consistent within $0.9 \sigma$. Right: Compared to SE systems, SubE systems are more diverse in planetary mass at the level of $\sim2.5\sigma$ significance. Taking our 16 NEA volatile-rich systems as a representative sample of systems containing only sub-Neptunes (SNs), we find that the mass uniformity of SE-only and SN-only systems are consistent within $0.9 \sigma$}
  \label{sub_earth}
\end{figure*}

Across our 48 CKS rocky systems, we identify 28 systems that contain only super-Earths ($1 R_{\oplus} \leq R_{p} \leq 1.6 R_{\oplus}$), and 8 systems that contain only sub-Earths ($R_{p} < 1 R_{\oplus}$). Carrying forward our CKS and NEA volatile-rich samples as representatives of the sub-Neptune population, we compare the $\mathcal{G}_{R}$ and $\mathcal{G}_{M}$  distributions of all three system types in Figure \ref{sub_earth}.

With regard to size uniformity, we find that the super-Earths and sub-Earths systems exhibit respective mean Gini values of $\overline{\mathcal{G}_{R, SE}} = 0.052_{-0.008}^{+0.010}$ and $\overline{\mathcal{G}_{R, SubE}} = 0.092_{-0.014}^{+0.031}$. We thereby observe that the sub-Earth systems exhibit greater size diversity than the super-Earth systems at the level of $2.05\sigma$ significance, and that the intra-system mass uniformity of the super-Earth sample is consistent with that of the sub-Neptune sample within $0.9 \sigma$.

We obtain very similar results for mass uniformity, where the super-Earth and sub-Earth systems exhibit respective mean Gini values of $\overline{\mathcal{G}_{M, SE}} = 0.184_{-0.035}^{+0.027}$ and $\overline{\mathcal{G}_{M, SubE}} = 0.302_{-0.102}^{+0.045}$, indicating that the super-Earth systems exhibit greater mass uniformity than the super-Earth systems with $2.51\sigma$ significance, but are consistent with the intra-system mass uniformity of sub-Neptunes systems within $0.9 \sigma$.

We therefore find that the increased mass diversity exhibited by our 48 CKS rocky systems is driven by the presence of sub-Earth, while their enhanced size uniformity is driven by the presence of super-Earths. We verify in Section \ref{sec:missing_plan} that the latter result is not an artifact of detection bias resulting from CKS rocky systems with unobserved sub-Earth members.

In terms of spacing uniformity, the 13 high-$N$ CKS rocky systems contain 5 super-Earth systems and 1 sub-Earth system. These sample sizes are far too small to perform the spacing uniformity test detailed in Section \ref{sec:spacing_res}.

The pronounced mass and size diversity exhibited by all systems containing sub-Earths may serve as mild evidence towards the breakdown of peas-in-a-pod architectures under 1 $R_{\oplus}$, though we caution that our findings are based off of smaller samples than our primary analyses, and that therefore nearly none of the validative tests in Appendix A may be applied. As such, we believe that further investigation of the influence of sub-Earths on intra-system uniformity warrants dedicated analyses of its own, and will likely benefit tremendously from the continued characterization and discovery of such diminutive worlds by ongoing and future observational campaigns.

\section{Discussion} \label{sec:discussion}
\subsection{Astrophysical Implications} \label{sec:theory}
We find that the dichotomy in intra-system mass uniformity presented thus yields broad qualitative consistency with several existing formation paradigms, a select few of which we shall briefly describe here. It has been remarked that local pressure maxima generated by early planet-disk interactions (\citealt{rice_bump}; \citealt{zhang_bump}; \citealt{dong}) may halt growth of existing planets and promote the accretion of companion cores (\citealt{carrera}; \citealt{xu_bump2}) to create a positive feedback loop amenable to the formation of a compact chain of planets all near the pebble isolation mass \citep{xu_bump}. Within such a framework, a dearth of material inside the snowline may stall planetary growth well below the isolation mass itself, and perhaps near or within the sub-Earth regime, thereby yielding greater mass diversity among planets lacking a significant volatile component. Alternatively, planet formation may be governed by the disruption of primordial, highly-uniform resonant chains (\citealt{terquem}; \citealt{morrison}; \citealt{broz}) by postnebular dynamical instabilities \citep{izidoro}, where the radius valley itself is shaped by a dichotomy between rocky and ice-rich planetary cores \citep{izidoro2}. In such a framework, systems containing these larger, ice-rich cores may exhibit a reduced propensity for giant impacts, allowing for greater maintenance of their primordial degree of mass uniformity. Alternatively, their rocky counterparts may experience more extreme collisional mass modulation, which may itself augment the available debris from which sub-Earth worlds may be born. Finally, a divergent evolutionary scheme may exist for which rocky planets interior to the snowline are formed in-situ with their mass scale determined by a series of giant impacts (\citealt{goldreich}; \citealt{emsenhuber}), while ex-situ worlds with surrounding vaporized hydrospheres experience substantial inward drift and a reduced propensity for collision, as well as a resistance to water loss from early giant impacts \citep{burn}.

Nonetheless, our demonstration of enhanced planetary mass uniformity for volatile-rich systems must lie in accord with our finding that such systems are also less uniform in planetary size compared to their rocky counterparts, especially those in the super-Earth regime. Although seemingly incongruous at a zeroth order level, the two trends may be reconciled intuitively based upon the highly mutable and diverse atmospheres of larger, volatile-rich worlds. Planetary mass is intimately linked to the formation history of a given system (e.g. \citealt{adams2}; \citealt{adams}), and is itself only significantly altered by large-scale, multi-body dynamical phenomena (e.g. \citealt{izidoro}; \citealt{goldberg}; \citealt{lammers}). Conversely, planetary size may be modulated to order unity by the growth or removal of a gaseous envelope that accounts for only a few percent of the total planetary mass (e.g. \citealt{owen_2013}), and while the removal of such atmospheres may indeed correlate strongly with prior interplanetary interactions \citep{ghosh2}, their presence and appearance are strongly governed by planet-dependent processes that lie distinct from system-level evolution. Examples of such phenomena include, but are not limited to: atmospheric erosion via core-powered processes (e.g. \citealt{ginzburg}; \citealt{berger_core}) or photoevaporation (e.g. \citealt{owen}; \citealt{mordasini}), inflation from tidal heating \citep{millholland_tides}, surface-atmosphere chemical reactions \citep{kite}, secondary outgassing \citep{kite2}, partial retention of a hydrogen envelope \citep{misener}, and mixing of an extended H/He envelope with steam \citep{burn}. Each of these processes, or any combination thereof, may thereby generate significant variations in individual planetary size while operating independently from the global formation mechanisms that promote mass uniformity for volatile-rich systems.

Compared to these trends in mass and size uniformity, we find that our evidence for enhanced spacing uniformity in rocky systems is not well-reproduced by existing evolutionary frameworks for multi-planet systems. Each of the paradigms thus discussed present comparatively quiescent and lower-entropy dynamical pathways for the assembly of volatile-rich worlds compared to their rocky counterparts, thereby collectively arguing for a greater uniformity in planetary spacing for the former in direct opposition to the trends presented in this work. It has been recently demonstrated that the presence of giant planetary companions can disrupt the spacing uniformity of the inner system \citep{he_gap}, though such a starkly preferential occurrence of these companions for volatile-rich systems cannot be accurately constrained with regard to our primary samples. As such, it remains to be determined if further analytical frameworks or simulations may successfully recover this surprising dichotomy in spacing uniformity.

\subsection{Comparison to the Solar System}
To contextualize our solar system in the broader context of the extrasolar subgiant population, we calculate the size, mass, and spacing Gini values for the four inner solar system planets, respectively obtaining values of $\mathcal{G}_{R} = 0.25$, $\mathcal{G}_{M} = 0.60$ and $\mathcal{G}_{\mathcal{P}} = 0.15$. 
The solar system values of $\mathcal{G}_{R}$ and $\mathcal{G}_{M}$ respectively lie at the 90th and 95th percentile of their analogous CKS rocky distribution, while the solar system value of $\mathcal{G}_{\mathcal{P}}$ exceeds the rocky distribution altogether and is markedly more consistent with the CKS volatile-rich systems. More amenable comparisons are found to CKS rocky systems containing only sub-Earths, as the solar system values of $\mathcal{G}_{R}$ and $\mathcal{G}_{M}$ respectively lie at the 75th and 33rd percentiles of the corresponding distributions for sub-Earth systems. We shall note as a crucial caveat that the efficacy of these comparisons is intrinsically limited by the residence of nearly the entire inner solar system beyond the $P = 100$ day cutoff within which all considered CKS systems are fully contained.
We thereby encourage caution in the interpretation of any first-order synergies or discrepancies in planetary uniformity emergent between the inner solar system and our samples of rocky extrasolar systems, as the former may be representative of a broader population that is slightly astrophysically distinct from the latter. Regardless, these architectural evaluations lend credence to the possibility of evolutionary divergences between the solar system and modal peas-in-a-pod architectures (e.g. \citealt{mishra_ii}), perhaps owing to inward migration or late instabilities of the solar system gas giants (\citealt{gomes}; \citealt{morbidelli_nice}; \citealt{tsiganis}; \citealt{walsh}), a relative dearth of solid material in the innermost regions of the protosolar disk (\citealt{kuchner}; \citealt{chiang_mmen}; \citealt{he_mmen}). The greater size diversity of the inner solar system compared to the typical sub-Earth system may also be caused by the more distant orbits of the former, as well as its wider orbital separations and greater interplanetary Hill spacing in comparison to the vast majority of \textit{Kepler} configurations \citep{weiss_rad}. Accordingly, the ongoing search for long-period giant companions to \textit{Kepler} multi-planet configurations \citep{weiss_kgps} may serve to provide more suitable comparative analogs for our solar system as a whole, which will further elucidate its evolutionary history in the context of of rocky extrasolar systems at large. The discovery of several terrestrial worlds, and possibly multi-planet configurations, with Earth-like orbits is an expected outcome of the upcoming PLAnetary Transits and Oscillations of stars mission (\citealt{plato}; \citealt{plato_earth}), and the observational characterization of their associated system architectures, specifically their inner edges \citep{batygin_edge}, may lend further credence to the notion that such configurations analogous to the inner solar system formed from postnebular collisional debris as opposed to a gas-rich primoridal disk \citep{qian}.

\section{Summary and Conclusions}
\label{conclusion}
In this work, we performed a direct statistical comparison of intra-system size, mass, and spacing uniformity for systems containing only small ($R_{p} \leq 1.6 R_{\oplus}$), rocky planets and those containing intermediate-size ($1.6 R_{\oplus} < R_{p} \leq 4 R_{\oplus}$), volatile-rich worlds. We shall enumerate here our primary results:

\begin{itemize}
    \item We compared the respective degrees of size uniformity for a sample of 48 CKS rocky systems and 118 CKS volatile-rich systems, finding that the former exhibits greater intra-system size uniformity with $\sim4.0\sigma$ significance.
    \item We inferred planetary masses for the 48 CKS rocky systems from the Earth-like (30\% Fe – 70\% MgSiO$_{3}$) $M$-$R$ relation from \citet{zeng}. We compared this sample to a collection of 16 volatile-rich systems with precise mass measurements from the NEA, and demonstrated that the rocky systems are more diverse in mass at the level of $\sim2.6\sigma$ significance.
    \item We identified from comparison of 13 high-$N$ ($N_{p} \geq 3$) CKS rocky systems and 30 high-$N$ volatile-rich systems that the former sample exhibits greater uniformity in interplanetary spacing with $\sim3.0\sigma$ significance.
    \item We isolated from our CKS rocky sample 28 systems containing only super-Earths ($1 R_{\oplus} \leq R_{p} \leq 1.6 R_{\oplus}$) and 8 systems containing only sub-Earths ($R_{\oplus} < 1 R_{\oplus}$), and demonstrate that the latter sample is more diverse in both planetary size ($\sim2.1\sigma$) and mass ($\sim2.5\sigma$). This suggests that the enhanced size uniformity of the CKS rocky systems is driven by the presence of super-Earths, while its enhanced mass uniformity is driven by the presence of sub-Earths.
    \item In Appendix \ref{sec:bias}, we perform several validative tests for our primary size, mass, and spacing uniformity results, finding that they are not significantly impacted by sample size effects, detection biases, varied assumptions of planetary bulk composition, and confounding astrophysical effects from the presence of near-resonant systems or low-mass hosts.
\end{itemize}

\section*{Acknowledgments} \label{sec:acknowledgments}
The crux of this paper, which employs Earth's composition as a basis to accurately determine the mass of small planets, draws inspiration from the work of \citet{xu_bump} and has been confirmed through valuable discussions with Fei Dai. We wish to extend our fullest gratitude to the anonymous referee for their detailed and extremely insightful feedback on the initial version of our manuscript, which led to the substantial strengthening and restructuring of our work into its current form. We would also like to thank the anonymous statistics editor for their valuable commentary regarding the general framework of our analysis, as well as Darin Ragozzine for overseeing the scientific review of our work. We are tremendously appreciative of the feedback and discourse provided by Konstantin Batygin, whose insights directly motivated our considerations of the sub-Earth population. Additionally, we extend our gratitude to Cristobal Petrovich and Jiwei Xie for their profound insights on contextualizing the properties of the inner solar system within our analysis. We also thank Brandon Radzom, Jessica Ranshaw, Xian-Yu Wang, and Emma Dugan for their support during the preparation of this manuscript. S.W. gratefully acknowledges the support received from the Heising-Simons Foundation through Grant No. 2023-4050.

\bibliographystyle{aasjournal}
\bibliography{main}

\clearpage
\appendix
\section{Evaluation of Biases} \label{sec:bias}
\subsection{Compositional Assumptions}\label{sec:comp_assume}
Despite the substantial efficacy of the \citet{zeng} Earth-like (30\% Fe – 70\% MgSiO$_{3}$) model in describing the $M-R$ behavior of small, rocky worlds (\citealt{rodriguez_martinez}; \citealt{rubenzahl}), we wish to ascertain the effects of alternative compositional assumptions on primary statistical results. We repeat our mass uniformity test (\ref{sec:mass_result}) with two additional semi-empirical $M$-$R$ prescriptions from \citet{zeng}, a more iron-rich (50\% Fe – 50\% MgSiO$_{3}$) compositional model and a “pure rock” model that are respectively consistent with slightly higher and lower bulk densities than the Earth-like curve (see Figure \ref{MR}). We find that the CKS rocky systems still exhibit greater mass diversity at respective significance levels of $2.53 \sigma$ and $2.51 \sigma$.

To account for a greater degree of variation in planetary composition both across the entire small planet population and within individual systems, we also perform a 1000 iterations of our mass uniformity test where the composition of each of the 118 CKS rocky planets is individually assigned at random from the three compositional models (Earth-like, iron-rich, pure rock) considered here. We find from this procedure that rocky systems exhibit a $2.66 \sigma$ preference for greater intra-system mass diversity. This increase in statistical significance from our primary $2.61 \sigma$ result agrees well with intuition: given that the CKS rocky systems are more diverse in planetary mass even when all of these masses are inferred directly from a singular, compositionally-uniform $M$-$R$ relation, it follows that this diversity shall only increase if individual planets are each subject to a broader set of compositionalassumptions and corresponding $M$-$R$ profiles. As such, we find that our primary $2.61 \sigma$ mass uniformity result is likely a lower limit on the true level of discrepancy between rocky and volatile-rich systems.

\subsection{Sample Size Effects}\label{sec:samp_size}
We shall consider briefly here the extent to which sample size effects may inform the results derived from our comparative statistical analyses. We thus consider 1000 iterations of each comparative procedure where systems are not partitioned by compositional classification: for each of the three planetary characteristics of interest (size, mass, spacing), we concatenate the relevant rocky ($N_{sys,rock}$) and volatile-rich ($N_{sys,vol}$) samples to form a mixed set of ($N_{sys,rock} + N_{sys,vol}$) total systems, and across 1000 testing iterations, we draw from this mixed set a random subsample of systems with size $\min\{N_{sys,rock}, N_{sys,vol}\}$ to which we apply, along with the complementary random mixed subsample of $\max\{N_{sys,rock}, N_{sys,vol}\}$ systems, the relevant comparative testing procedure.

For considerations of size uniformity (CKS rocky and CKS volatile-rich), mass uniformity (CKS rocky and NEA volatile-rich), and spacing uniformity (high-$N$ CKS rocky and high-$N$ CKS volatile-rich), we obtain respective significance distributions of $-0.00 \sigma \pm 1.66 \sigma$, $0.03 \sigma \pm 1.20 \sigma$, $-0.04 \sigma \pm 1.71 \sigma$. These distributions confirm the expectation of a null result for a compositionally agnostic procedure, and indicating that our primary $4.04 \sigma$, $2.61 \sigma$, and $3.01 \sigma$ results respectively hold a $< 1$\%, $< 2$\%, and $< 4$\% probability of coincidental occurrence from sample size effects.

\subsection{Presence of Systems with Planets Straddling the Compositional Dichotomy}
\label{sec:overlap}
Since the compositional partitioning scheme utilized in this work is itself predicated upon a discrete planetary size cutoff of $R_{cut} = 1.6 R_{\oplus}$, we wish to investigate the influence of planets whose radii overlap this boundary within their $1\sigma$ measurement uncertainties. For each sample considered in this work, we remove any systems containing at least one such planet, resulting in the omission of 6 of 48 CKS rocky systems (including 2 of 13 high-$N$ systems), 22 of 118 CKS volatile-rich systems (including 6 of 30 high-$N$ systems), and 3 of 16 NEA volatile-rich systems. From these revised samples, we find that, compared to their volatile-rich counterparts, rocky systems are more uniform in size, less uniform in mass, and more uniform in spacing at respective significance levels of $1.76\sigma$, $2.41\sigma$, and $2.88 \sigma$. We note that the relatively large drop in significance for our size uniformity result may be intuitively attributed to the high degree of compositional and physical diversity exhibited by volatile-rich planets lying close to the $R_{cut} = 1.6 R_{\oplus}$ boundary, as such worlds may harbor their volatile component predominantly in the form of an extremely tenuous H/He atmosphere \citep{misener}, a high-pressure H$_{2}$O layer \citep{kite3}, or ices sequestered in the planetary core (\citealt{izidoro2}; \citealt{burn}), in contrast to the massive H/He envelopes commensurate with larger worlds (\citealt{owen_2013}; \citealt{fulton_gap}).

\subsection{Presence of Near-Resonant Systems}\label{sec:near_res}
Given that planetary systems containing at least one pair close to first-order MMRs exhibit enhanced size uniformity compared to systems that are are entirely nonresonant \citep{goyal2}, it remains to be seen if such effects may confound any emergent discrepancies in planetary size or mass uniformity between different compositional regimes. Following the methodology of \citet{goyal2} we search for planetary pairs in close proximity to first-order MMRs using the $\zeta_{2,1}$ parameter from \citet{fab}:

\begin{equation}
\label{eq2}
    \zeta_{2,1} = 3\left( \frac{1}{\mathcal{P}-1} - \text{Round}\left[ \frac{1}{\mathcal{P}-1} \right] \right),
\end{equation}

where any system containing at least one pair with $|\zeta_{2,1}| \leq 0.25$ is regarded as near-resonant. 

We find that 4 out of 16 systems in the NEA volatile-rich sample (25\%) and 11 out of 48 systems in the CKS rocky sample (22.9\%) are near-resonant, qualitatively indicating that our mass uniformity result may not be influenced appreciably by an imbalance of near-resonant configurations. However, we find that only 16 of the 118 systems in the CKS volatile-rich sample (13.6\%) are similarly close to MMR, possibly indicating that the enhanced size uniformity of the CKS rocky sample may be generated by a greater occurence of high-uniformity, near-resonant pairs in the latter group. To investigate this claim with greater rigor, we repeat our size and mass uniformity tests for only the 37 CKS rocky, 102 CKS volatile-rich, and 12 NEA volatile-rich systems that that are completely devoid of near-resonant planetary pairs. We find that rocky systems exhibit greater size uniformity and greater mass diversity at respective significance levels of $3.54\sigma$ and $2.67\sigma$, affirming that our primary statistical results for these characteristics are not confounded by an imbalance in the prevalence of near-resonant configurations within the relevant samples. 

\subsection{Presence of Low-Mass Host Stars}\label{sec:fgk}
M-dwarfs exhibit an overabundance of close-in super-Earths and a dearth of gas giants compared to main sequence stars (\citealt{dressing}; \citealt{mulders}; \citealt{sabotta}; \citealt{ment}), which are likely indicative of distinct planetary formation dynamics. Additionally, planetary mass uniformity has been observed to scale with the planet-to-star mass ratio as opposed to absolute planetary mass \citep{xu_bump}, suggesting that low-mass stars may harbor highly uniform planetary chains despite their relative lack of solid disk mass \citep{lin}. We thereby repeat our comparative analyses with all relevant samples restricted to FGK main-sequence hosts ($4700 \leq T_{eff} \leq 6500$, $\log g \geq 4$). This constraint leaves 9 of 16 NEA volatile-rich systems, all 48 CKS rocky systems, 103 of 118 CKS volatile-rich systems, and 29 of 30 CKS high-$N$ volatile-rich systems, from which our size, mass, and spacing uniformity significance results are respectively maintained at levels of $4.04\sigma$, 2.60$\sigma$, and 3.11$\sigma$, though we note the potential influence of small-sample effects on the mass uniformity result.

\subsection{TTV vs. RV Mass Measurements}\label{sec:ttv_rv}
At a fixed planetary size, RV-based searches consistently favor the detection of more massive objects than the TTV measurements (\citealt{weiss2014}; \citealt{hadden}), and it thus remains to be determined the extent to which the presence of both measurement techniques in our NEA volatile-rich sample may impact the primary statistical results presented in this work. Within our NEA volatile-rich sample, we find that 4 of the 37 total planets have their masses constrained via TTV fitting: KOI-1599.01, KOI-1599.02, Kepler-1705 b, and Kepler-1705 c (\citealt{panichi}; \citealt{leleu3}). As the TTV mass measurements for these objects all lie below the median of the distribution yielded by the remaining 35 RV mass measurements, we subsequently remove the two associated systems from our volatile-rich sample and compare the remaining 14 systems to our CKS rocky sample via our mass uniformity test. We find that even with the limitation of our data to RV mass measurements alone, the CKS rocky systems exhibit greater mass diversity than the NEA volatile-rich systems at the level of $2.53 \sigma$ significance, thereby confirming that the consideration of both TTV and RV measurements do not appreciably inform our primary statistical results.

\subsection{Detection Bias and Missing Planets}\label{sec:missing_plan}
The high-$N$ CKS rocky and volatile-rich systems boast vastly different distributions for their orbital periods and interplanetary spacing, with the former being exclusively close-in and compact ($P \lesssim 35$ days, $\mathcal{P} \lesssim 2.3$) while pairs in the latter exhibit much wider gaps ($\sim 27$\% of pairs with $\mathcal{P} \gtrsim 2.3$). As such, it remains to be seen if undetected external companions in the CKS rocky systems or missing planets in the gaps of the CKS volatile-rich systems impart any substantial bias on the size and spacing uniformity results presented in this work. We assess these potential biases using the heuristic framework from \citet{millholland_edge}: consider the expression for the signal-to-noise ratio (S/N) of a hypothetical \textit{Kepler} planet, 

\begin{align}
    \text{S/N} &= \frac{(R_{p}/R_{\star})^{2}\sqrt{3.5\text{yr}/P}}{\text{CDPP}_{\text6h}\sqrt{\text{6h}/T}}\\
    T &=  13\text{hr} \left(\frac{P/\text{yr}}{\rho_{\star}/\rho_{\odot}}\right)^{1/3},
\end{align}

where $R_{p}$ and $R_{\star}$ are the solar and stellar radii, $\rho_{\star}/\rho_{\odot}$ is the solar-scaled stellar density, and CDPP$_{\text{6h}}$ is the combined differential photometric precision in the \textit{Kepler} light curve over 6 hours \citep{chirstiansen_cdpp}. If we assume for such a hypothetical planet a mock size $\widetilde{R_{p}}$ and the CKS minimum $(\text{S/N})_{min} = 10$, we may solve for its maximum period of detection to compare to our $P \leq 100$ day cut from Section \ref{sec:size_samp}:

\begin{equation}\label{pmax}
    P_{max}/\text{yr} = \left(\frac{91}{12}\right)^{3/2}\frac{(\widetilde{R_{p}}/R_{\star})^{6}}{\text{CDPP}_{\text6h}^{3} (\text{S}/\text{N})_{min}^{3}\sqrt{\rho_{\star}/\rho_{\odot}}}
\end{equation}

For our CKS rocky sample, we adopt for each system the radius of the outermost planet as $\widetilde{R_{p}}$, finding that 30 of 48 (6 of 13 high-$N$) systems may have an additional external companion within 100 days. For the CKS volatile-rich sample, we consider the median radius as $\widetilde{R_{p}}$, with 13 of 118 (4 of 30 high-$N$) systems harboring the possibility for an undetected intermediate member within 100 days. We repeat our size and spacing uniformity tests on the 23 CKS rocky (7 high-$N$) and 110 CKS volatile-rich (26 high-$N$) system that remain, finding that the rocky systems still exhibit enhanced size and spacing uniformity at respective levels of $3.49\sigma$ and $2.12\sigma$ confidence.

In Section \ref{sec:sub_earth}, we demonstrate that systems containing only sub-Earths exhibit a considerable degree of diversity in their constituent planetary sizes, indicating that the enhanced size uniformity observed for the CKS rocky sample as a whole is likely driven by the presence of super-Earths. However, it remains to be seen whether this phenomenon is itself a result of detection bias, such that rocky systems with a high degree of size uniformity may only appear as such due to undetected sub-Earth members. We investigate this notion directly by determining for each CKS rocky system the minimum radius ($R_{min}$) for which a hypothetical planet would have been detected by the CKS (S/N $= 10$) within the orbit of the outermost member ($P \leq P_{outer}$):

\begin{equation}
    R_{min} = \left(\frac{12}{91}\right)^{1/4}\left(\frac{R_{\star}}{R_{\odot}}\right)
    \left(\frac{\rho_{\star}}{\rho_{\odot}}\right)^{1/12}
    \sqrt{{\text{CDPP}_{\text6h}}\times (\text{S}/\text{N})_{min}}
\end{equation}

Given that 13 of the 48 CKS rocky systems have $R_{min} \geq 1 R_{\oplus}$, we compare the remaining 35 systems to our 118 CKS volatile-rich systems and find that the former still exhibit enhanced size uniformity at the $2.85 \sigma$ level. This discrepancy is further maintained at the $2.3 \sigma$ level for the exclusion of CKS rocky systems with $R_{min} \geq 0.9 R_{\oplus}$. These results suggest that the size uniformity discrepancy observed between rocky and volatile-rich systems is likely not driven by biased rocky systems with undetected sub-Earth companions, though the possible influence of hidden sub-Earths exterior to the currently known bounds of these systems warrants future investigation.



\end{document}